\documentclass[10pt,preprint]{article}
\usepackage[footskip=0.5in]{geometry}

\usepackage{dsfont}

\usepackage{graphicx}

\usepackage{amsmath,amssymb}

\usepackage{changepage}

\usepackage[utf8x]{inputenc}

\usepackage{textcomp,marvosym}

\usepackage{cite}

\usepackage{nameref,hyperref}

\usepackage{microtype}
\DisableLigatures[f]{encoding = *, family = * }

\usepackage[table]{xcolor}

\usepackage{array}

\newcolumntype{+}{!{\vrule width 2pt}}

\newlength\savedwidth

\newcommand\thickhline{\noalign{\global\savedwidth\arrayrulewidth\global\arrayrulewidth 2pt}%
\hline
\noalign{\global\arrayrulewidth\savedwidth}}

\makeatletter
\renewcommand{\@biblabel}[1]{\quad#1.}
\makeatother

\date{}

\usepackage{epstopdf}

\usepackage{multirow}

\begin{document}
\vspace*{0.2in}

 \begin{center}
{\Large
\textbf\newline{\textbf{Emerging interdependence between stock values during financial crashes}}
}
\end{center}
\begin{center}
Jacopo Rocchi\textsuperscript{1}, 
Enoch Yan Lok Tsui\textsuperscript{2}, 
David Saad\textsuperscript{1} 
\\
\bigskip
\textbf{1} Nonlinearity and Complexity Research Group, Aston University, Birmingham, B4 7ET, United Kingdom
\\
\textbf{2} Department of Physics, The Hong Kong University of Science and Technology, Clear Water Bay, Kowloon, Hong Kong, China
\\
\bigskip

%
%






*j.rocchi@aston.ac.uk
\end{center}
\section*{Abstract}

To identify emerging interdependencies between traded stocks we investigate the behavior of the stocks of FTSE 100 companies in the period 2000-2015, by looking at daily stock values. Exploiting the power of information theoretical measures to extract direct influences between multiple time series, we compute the information flow across stock values to identify several different regimes. While small information flows is detected in most of the period, a dramatically different situation occurs in the proximity of global financial crises, where stock values exhibit strong and substantial interdependence for a prolonged period. This behavior is consistent with what one would generally expect from a complex system near criticality in physical systems, showing the long lasting effects of crashes on stock markets.



\section*{Introduction}

Financial markets in general and the drivers for market crashes in particular have been intensively investigated by the physics community in recent years. The interest in markets' behavior stems from a number of different reasons relating both to the narrow financial interests of shareholders and investors and, arguably more importantly, to the devastating impact financial turmoil may have on national economies, leading to harsh social consequences and societal unrest.
Moreover, the observation of a sudden and dramatic collapse in complex systems intrigues the scientific community due to its resemblance to collective rearrangement in physical systems at critical points.
This problem is at the core of the emerging field of econophysics~\cite{mantegna2000introduction, plerou2000econophysics, bouchaud2003theory}, which borrows mathematical and physical tools such as random matrix theory~\cite{plerou1999universal, laloux2000random}, clustering analysis~\cite{fenn2011temporal}, extreme and rare events~\cite{sornette2012dragon}, agent based models~\cite{gualdi2015tipping} and network theory~\cite{mantegna1999hierarchical}, to name a few, to tackle the complexity of economical and financial systems.
Network theory provided a framework for analyzing economic structures~\cite{mantegna1999hierarchical, tumminello2005tool, allen2008networks, haldane2013rethinking} from the perspective of complex systems~\cite{strogatz2001exploring}, and is rooted in a much earlier search for structures in financial markets~\cite{elton1971improved}.
The success of this approach is partly due to shortcomings of existing economics-based theories in addressing the complexity of financial systems, and partly because many phenomena such as financial bubbles, herding, contagion and crashes found a natural interpretation in physical models that involve multi agents, collective behavior, influence spreading and phase transitions.

Phase transitions are sudden reorganizations of the system occurring when an external parameter, such as the temperature, is tuned to a critical value.
Moreover, at this critical point, the system is scale-invariant, leading to a power law behavior of the observables whose critical exponents may be studied using renormalization group and scaling theory~\cite{wilson1971renormalization}.
A similar symmetry, discrete scale invariance~\cite{sornette1998discrete}, has been shown to give rise to log-periodicity of prices in the proximity of financial collapses~\cite{sornette2001significance}.
Important contributions to the analogy between financial crashes and phase transition came also from the study of ecological and climate systems. In particular, a large effort has been devoted to studying the precursors of collapses~\cite{dakos2008slowing, scheffer2009early, scheffer2012anticipating}, showing that crashes are usually anticipated by a loss of resilience in the system; in other word, when approaching a critical point, perturbations take more time to be reabsorbed and are more likely to propagate.
This picture is consistent with that of physical systems near the critical point of a phase transition, where the cross correlation between the fundamental degrees of freedom of the systems is very large and the system exhibits high susceptibility.

The study of correlations in financial networks has started a couple of decades ago and led to the development of several effective algorithms for extracting the underlying network topology~\cite{mantegna1999hierarchical, bonanno2003topology, onnela2003dynamics, bonanno2004networks, tumminello2005tool, tumminello2007correlation}. Their hierarchical structure could then be used to identify groups of stocks in terms of the corresponding economical sectors. These works originally focused on same-time correlations and only recently have been generalized to deal with the concept of causality~\cite{granger1988some} in financial data~\cite{curme2015emergence}.
The study of directed influences, aiming at predicting future prices, is interesting for investors for maximizing their returns; however, it also has the potential to forecast the macroscopic behavior of markets.
This objective is highly ambitious, but predicting the behavior of macroscopic properties from observation and modeling is at the heart of statistical physics and is of theoretical interest on its own right.
Attempts in this direction has been made using Granger causality~\cite{huth2014high} and information theoretical methods such as mutual information and Transfer Entropy (TE)~\cite{schreiber2000measuring}, which outperform simple retarded correlations in capturing non-linear influences~\cite{fiedor2014information,fiedor2015}.

In this paper we generalize existing information theoretic approaches to study temporal interdependencies between financial indices in a time period of several years. More precisely, we analyze interdependencies between stocks of the FTSE 100 companies, which includes the 100 largest companies (in terms of market capitalization) listed on the London Stock Exchange, from 2000 to 2015.
Exploiting the physical intuition that financial crashes may be anticipated by periods of large susceptibility, we use our method to investigate directed influences among the corresponding index constituents looking for a similar behavior, with the limited resolution of daily stock values.
Most of the literature in this area relies on intra-days influences since it is commonly believed that traders are well informed when making their decision, so that directed correlations and influences can be detected only at short time scales. This belief is an interpretation of the ``efficient market hypothesis"~\cite{fama1965behavior}, according to which prices reflect all the available information and thus there is no hope to predict and outperform the market.
While this is true to some extent, other studies showed that this hypothesis is too simplistic and that daily stock prices do not behave as trivial random walks~\cite{bossomaier2013information, fiedor2014frequency, fiedor2014information}.
By analyzing moving periods of roughly two years each, our findings support the common view that little information can be extracted from the past to predict future values at daily time scales. However, we notice that crisis periods make an exception to this rule. In fact, measuring the overall information flow between index components we detect strong interdependencies for periods corresponding to the crash of 2008 and the Eurozone debt crisis of 2010-2012.

The paper is organized in the following way. The second section focuses on the materials and methods used, and is divided to three subsections. The first introduces the information theoretical methods which are commonly used for extracting direct influences, the second provides details on the dataset and the third explains the null model used to validate our analysis.
The third section contains the results and the following ones provide a discussion and conclusion. Further details on the methods are provided in the Supporting Information.

\section*{Materials and Methods}

\subsection*{Measuring influences}

In order to measure the influences among stocks we used an Information theoretical tool which follows from a generalization of the Symbolic Transfer Entropy (STE)~\cite{staniek2008symbolic}.
TE and STE are powerful methods able to measure the amount of information flow between time series and thus can be used to reconstruct the network of influences between components of a complex systems.
Some of the most interesting cases where they have been successfully employed include network reconstruction of functional areas of the brain~\cite{rubinov2010complex,vicente2011transfer}, the study of  social phenomena~\cite{borge2016dynamics} and the influence of social networks on financial markets~\cite{souza2016nonlinear}.
They have also been used in finance to analyze the relations between indices~\cite{marschinski2002analysing, kwon2008information} and components of indices~\cite{fiedor2015}.

Transfer entropy evaluates the information gained on future values of a time series $X(t)$ by observing past values of another time series $Y(t)$ in addition to the past values of $X(t)$, relying on estimating the probabilities of occurrences of time series values. For real valued time series this is more difficult but STE, making use of symbolization, provides an effective solution to this problem.
A symbol of $k-$literals of the time series $X(t)$ at time $s$ is obtained by reordering the last $k$ values of the time series at time $s$ (i.e. $\{x_{s-k}, \ldots, x_{s-1} \}$) in an ascending order.
By generalizing the time step from $1$ to an integer $\delta$ we transform the data to a set of $k-$dimensional symbols at times $s+\delta$, denoted by $\hat{x}^{k}_{s+\delta}$. The role of the time scale $\delta$, which reflects the inherent effective delayed interaction between components will be investigated later.
A more formal definition as well as further details on the corresponding information theoretical measures are provided in the Supporting Information.

In this work, we evaluate the influence of time series $Y(t)$ on $X(t)$ by computing the following quantity
\begin{equation}
T_{Y\rightarrow X} = \sum_{\hat{x}^{k+1}_{t+\delta}, \hat{x}^{k}_t, \hat{y}^{k}_t } p\left(\hat{x}^{k+1}_{t+\delta}, \hat{x}^{k}_t, \hat{y}^{k}_t \right) \log_2 \left( \frac{p\left(\hat{x}^{k+1}_{t+\delta} | \hat{x}^{k}_t, \hat{y}^{k}_t \right)}{p\left(\hat{x}^{k+1}_{t+\delta} | \hat{x}^{k}_t \right)}  \right)\:.
\label{eq:tildeT}
\end{equation}
This is the Kullback-Leibler divergence between the probabilities $p\left(\hat{x}^{k+1}_{t+\delta} | \hat{x}^{k}_t, \hat{y}^{k}_t \right)$ and $p\left(\hat{x}^{k+1}_{t+\delta} | \hat{x}^{k}_t \right)$, averaged over past symbols. Alternatively it can be viewed as the difference between the conditional entropies of the two probabilities.
If $Y(t)$ contains no information about $X(t)$ this measure is zero. Practically, due to the noisy nature of the data this never happens and one finds non zero values even when the two systems do not interact. This calls for the introduction of a null model to extract the genuine underlying behavior from the dataset; this will be discussed in detail later on. Finally we would like emphasize that our measure aims at predicting $(k+1)$-dimensional symbols by looking at $k$-dimensional historic symbols as explained in the Supporting Information.

\subsection*{Dataset}

\begin{figure}[ht]
\centering
\includegraphics[width=130mm]{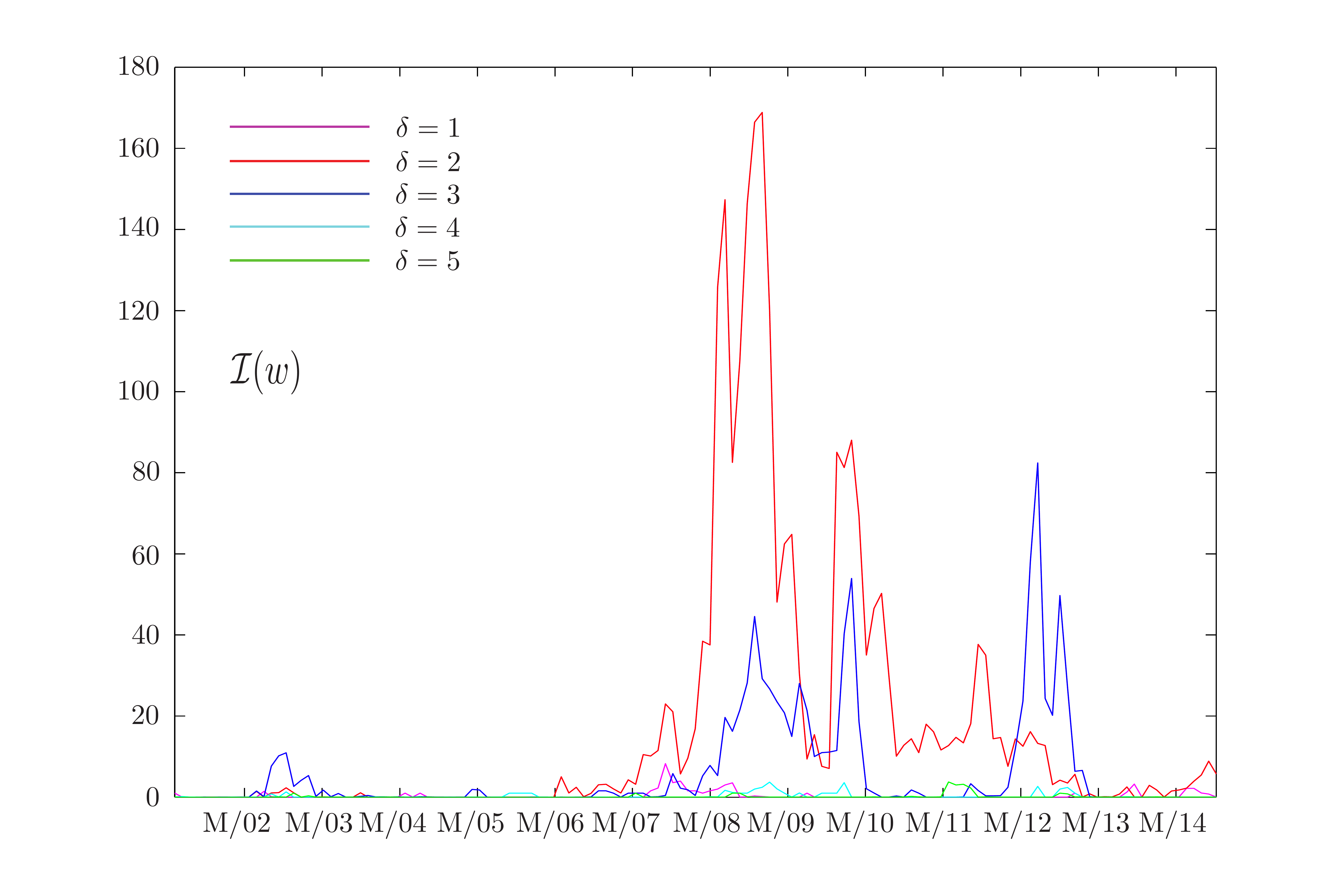}
\label{Fig1}
\caption{The behavior of the total information flow $\mathcal{I}(w)$, defined in eq. (\ref{eq:totalIw}), at different time scales $\delta$. Each time window $w$ is $500$ days long. The date associated to each $w$ is the middle of the time window considered. The $x$-axis tick marks represent the first of March of every year. While at short time scales (less than 3 days) we observe a peak around the two major financial crises of the last decades, this effect fades away as $\delta$ increases. Interestingly, the results at $\delta=2$ carries much more information than those at $\delta=1$.}
\end{figure}

\begin{figure}[ht]
\centering
\includegraphics[width=130mm]{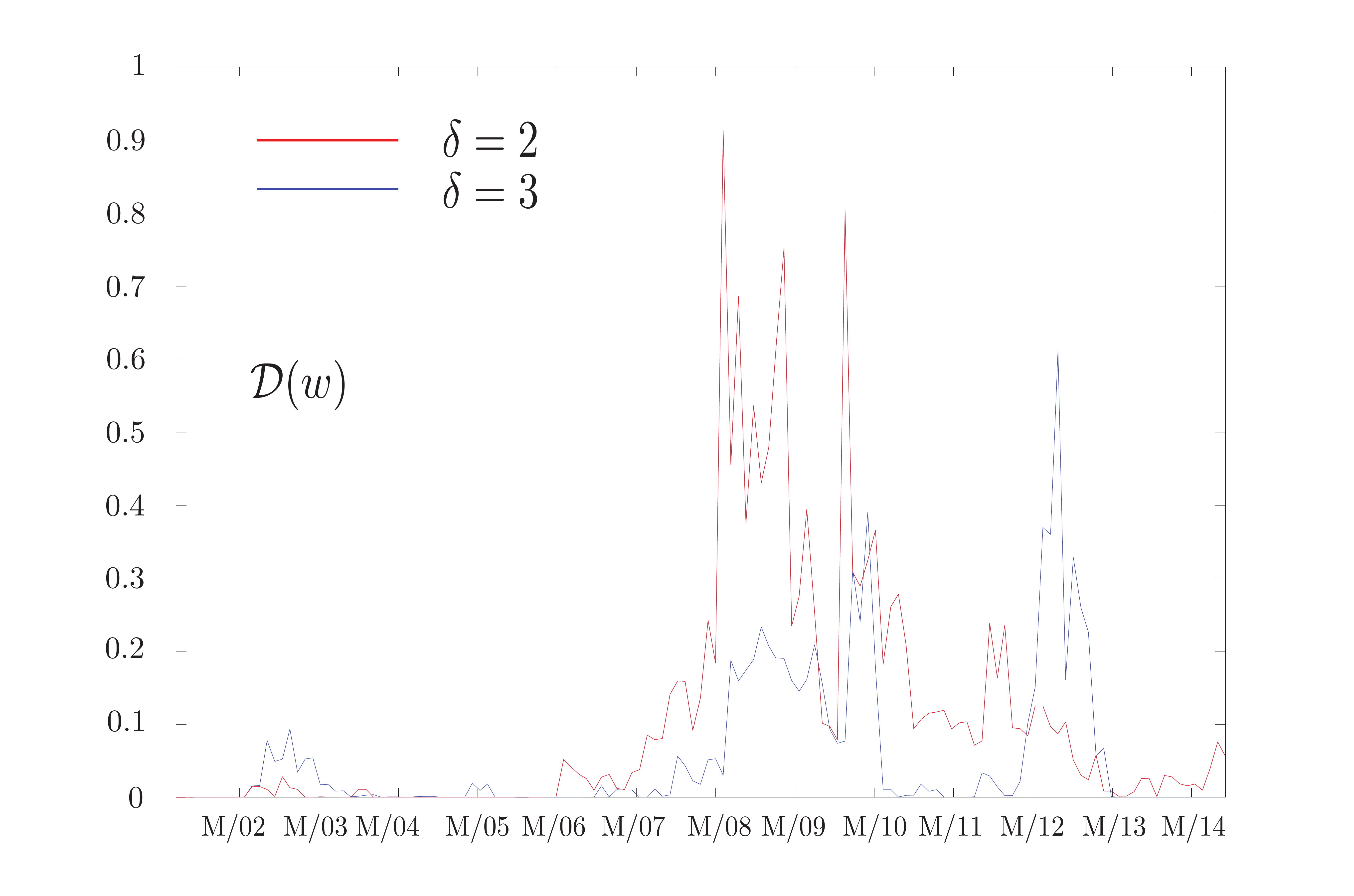}
\label{Fig2}
\caption{Plot of $\mathcal{D}(w)$, the information flow in two consecutive time windows, defined in eq. (\ref{eq:eqDdiff}), at time differences $\delta=2$ and $\delta=3$. Each time window $w$ is $500$ days long. The date associated to each $w$ is the middle of the time window considered. The $x$-axis tick marks represent the first of March of every year.
This quantity measures the evolution of the detected structure of influences. We observe a smooth behavior, meaning that structures in consecutive time windows are similar, except for during crises where more pronounced market readjustment take place.}
\end{figure}

\begin{figure}[ht]
\centering
\includegraphics[width=130mm]{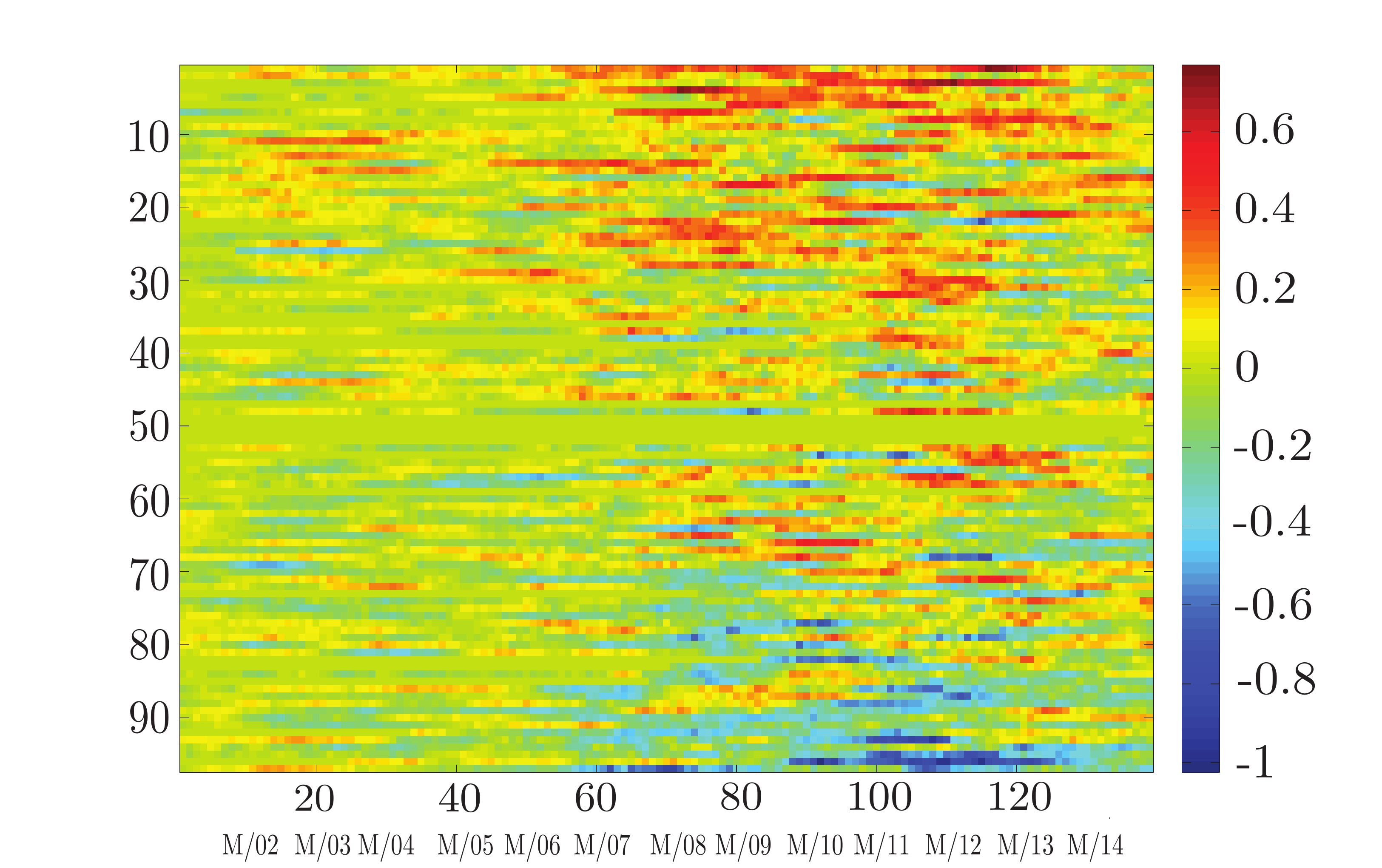}
\label{Fig3}
\caption{For each component $n$, we evaluate the information directionality flow $\Delta_n$, defined in eq. (\ref{eq:Delta_n}), measuring how much the component has influenced (or has been influenced by) the market. Positive values are associated to lead effects. The horizontal axis refers to the window time index $w$. The vertical axis refers to the component index, ordered according to the value of the overall influence over time. It is interesting to see how strength and directionality of influences become clearer and more emphasized at time of financial crises. A closer look at these values is provided in Figs.~4 and 5.}
\end{figure}

\begin{figure}[ht]
\centering
\includegraphics[width=130mm]{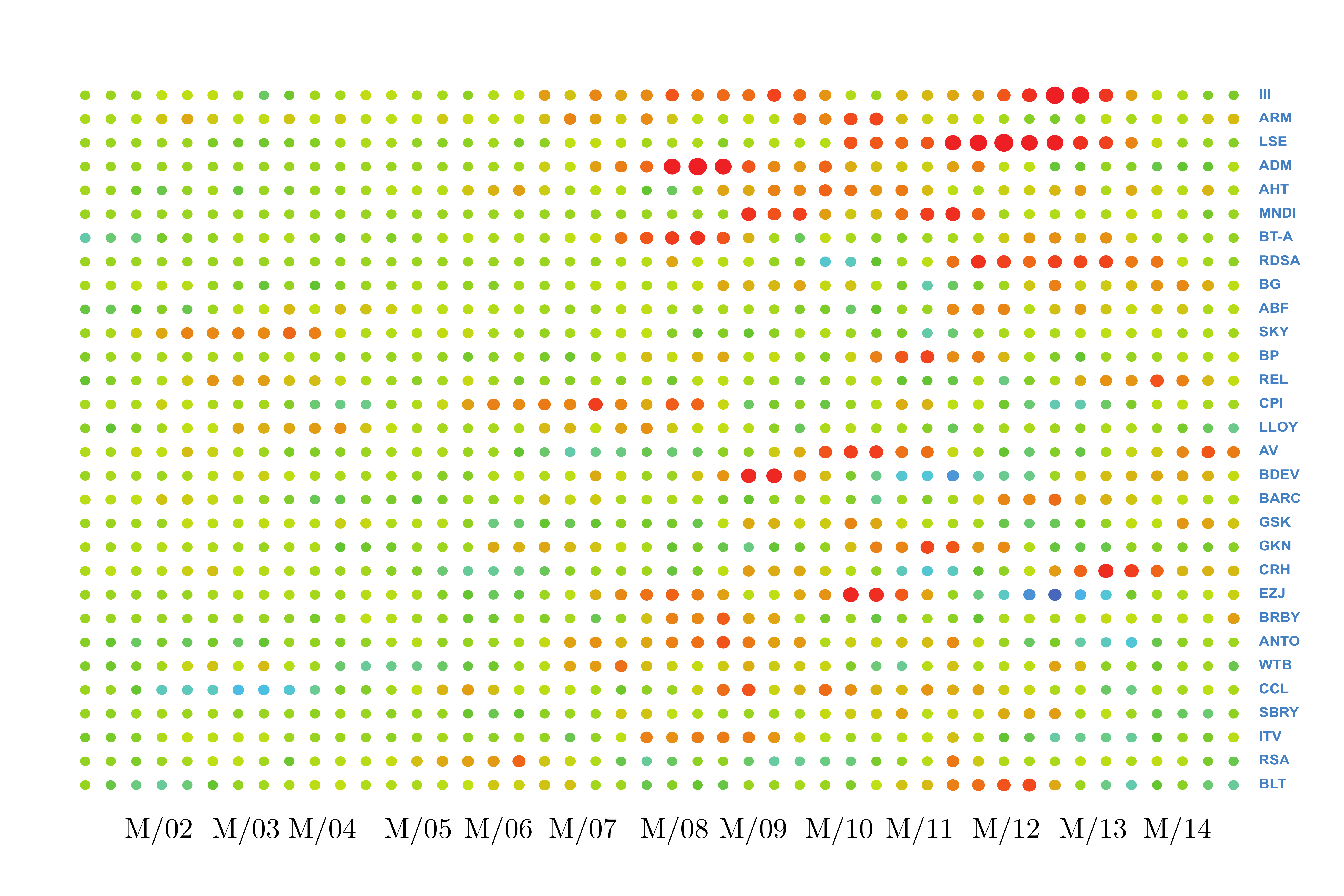}
\label{Fig4}
\caption{To identify more clearly stocks led by the market, we present the information of Fig.~3, but focussing on the $30$ components with the \emph{largest} directionality flow values.
For the sake of clarity each time tick has been obtained by averaging three consecutive time windows. So we have about $45$ different ticks rather than the original $140$ time windows.}
\end{figure}

\begin{figure}[ht]
\centering
\includegraphics[width=130mm]{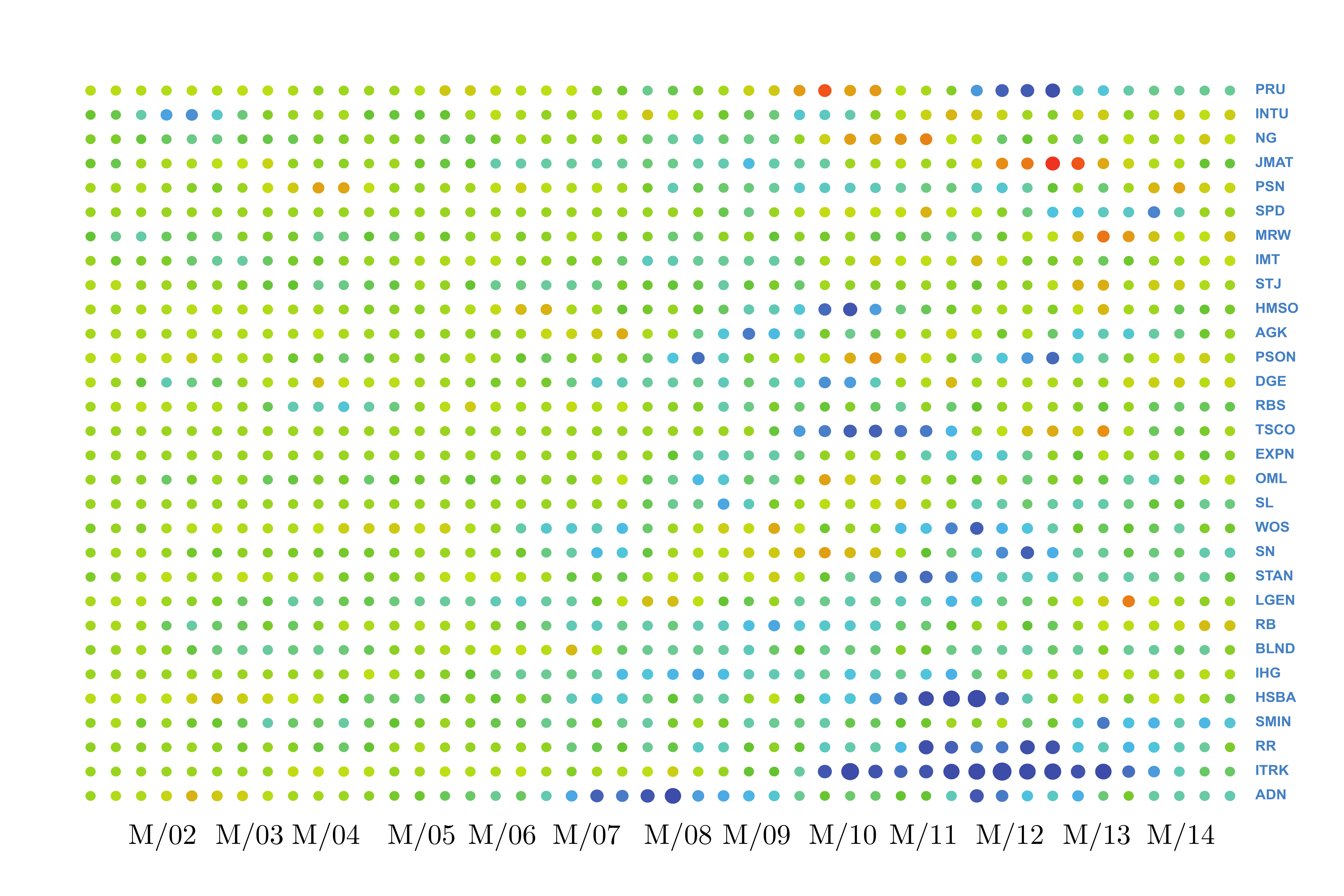}
\label{Fig5}
\caption{To identify more clearly stocks led by the market, we present the information of Fig.~3, but focussing on the $30$ components with the \emph{smallest} directionality flow values.
For the sake of clarity each time tick has been obtained by averaging three consecutive time windows. So we have about $45$ different ticks rather than the original $140$ time windows.}
\end{figure}

We collected financial time series data of the component stocks of FTSE 100 from 3 January, 2000 to 15 May, 2015 (around $4000$ trading days), available from Yahoo! Finance~\cite{YahooFin}. Discarding short-lived stocks, we labeled the remaining $N=97$ stocks from $1$ to $97$ in ascending alphabetical order of their ticker symbol. These time series have a time resolution of one (trading) day and we focused on the closing prices. Instead of looking at the whole time series, we analyze time windows of $\Omega = 500$ days. This time window is being shifted by $\omega=25$ days across about $4000$ trading days for which we have data, investigating the time evolution of the network structure.
Since the stationarity hypothesis, useful in estimating the probabilities~(\ref{eq:tildeT}), is unlikely to hold for long periods, studying shorter time windows would help in getting more reliable estimates.
We process the time series linked to each stock in order to obtain the geometric returns $r(t)$ at the time scale $\delta$:
\begin{equation}
r(t)=\log [p(t+\delta)]-\log[p(t)]\:,
\label{eq:return}
\end{equation}
where $p(t)$ is the price at time $t$ (closing price on day $t$). In each time window, we compute the information flow between time series at different $\delta$ values using Eq.~(\ref{eq:tildeT}); while log-ratios with large $\delta$ values are expected to carry little or no information we will show that also log-ratios with small $\delta$ values do not, in certain periods.
To further control errors and ensure that the stocks considered had existed for long enough to give rise to meaningful influences, we restrict the computation of $T_{Y \rightarrow X}$ to cases where the number of days the considered pair of stocks $\{X(t), Y(t)\}$ have in common is at least $80\%$ of the time window $\Omega$.

\subsection*{Surrogate dataset}

To validate our results and eliminate spurious instances of entropy transfer, we construct a null model of non interacting components. This may be done in several ways~\cite{marschinski2002analysing, kwon2008information, fiedor2015, borge2016dynamics}, ranging from a random reshuffle of the original time series to more refined methods~\cite{theiler1996constrained,schreiber1998constrained}. A simple reshuffling of data, while clearly destroying the interdependence among different time series, also destroys the single time series structure. The null model we use, based on the theory of surrogate data~\cite{theiler1992testing}, does allow one to preserve the spectral properties of the original spectrum in spite of the randomization. Under the assumption that the single time series structure can be effectively represented by the power spectrum of the signal, a general time series $X(t)$ can be randomized via the generation of the time series
\begin{equation}
\tilde{X}(t)=\mathcal{F}^{-1}\left[ X(k) e^{i \phi(k)} \right]\:,
\end{equation}
where $X(k)$ is the Fourier transform of the original signal, $X(k) = \mathcal{F} [ X(t)]$, and $\phi(k)$ is a random phase attached to each Fourier component such that $\phi(-k)=-\phi(k)$, so that $\tilde{X}(t)$ is real.
The series $\tilde{X}(t)$ is thus a randomized version of $X(t)$ but having the same power spectrum.
We construct the null model by randomizing the original time series of the closing prices. Then, for a given $\delta$, we process these time series to obtain random returns using Eq.~(\ref{eq:return}) and compute the influences between the surrogate time series using Eq.~(\ref{eq:tildeT}).
We compared the information flow in our original dataset and the null model to identify true information from noise since the latter does not contain genuine information flow between series.

\section*{Results}
\label{sec:sec_result}
We analyzed the evolution of the network of influences between stocks in each of the (about) $140$ time windows indexed by $w$.
We associate a value $I(X,Y) \in [0,1]$ to each directed link $\{X \rightarrow Y\}$ in order to measure the amount of genuine information flow from $X(t)$ to $Y(t)$.
This is done by comparing the measure $T_{X \rightarrow Y}$ with the corresponding quantities computed for the surrogate dataset; this is used to estimate the probability that information flow values obtained in the real dataset have been obtained at random. Further details are provided in the Supporting Information.

The quantities $I(X,Y)$ are supposed to vary slowly from one time window to the next, say $w$ to $w+1$; conversely, the parameter $\omega=25$ that controls the shift between consecutive time-windows, may be too large or our results may not be sufficiently stable with respect to small changes.
To address this issue we introduced the quantity
\begin{equation}
\mathcal{D}(w) = \frac{1}{N}\sum_{i=1}^N \left| \sum_{j=1}^N I_{w+1} (X_i, X_j) - I_w (X_i, X_j) \right|\: ,
\label{eq:eqDdiff}
\end{equation}
where, the expression within the absolute value sign is the change in the genuine information flow originating from stock $i$ in two consecutive time windows.

To estimate the \emph{total} information flow, we introduce the quantity
\begin{equation}
\mathcal{I}(w)=\sum_{X,Y} I(X,Y)\:,
\label{eq:totalIw}
\end{equation}
and study its behavior as a function of $\delta$ and $w$.
While the results for large $\delta$ confirm our expectations that no information flow can be detected, the results obtained at small $\delta$ as a function of $w$ are much more interesting.
In particular, analyzing geometric return time series by setting $k=2$ in Eq.~(\ref{eq:tildeT}), we obtain the results shown in Fig.~1.
At small values of $\delta$ this quantity sharply peaks around the financial crisis of 2008, while this effect fades away as $\delta$ increases.
We also notice that a similar behavior is observed during the period of the Eurozone debt crisis between 2010 and early 2013, and then again fades away as $\delta$ increases.
The robustness of our results can be checked analyzing the behavior of $\mathcal{D}(w)$. As can be seen in Fig.~2, this quantity, representing the information flow in two consecutive time windows, is close to zero most of the time and peaks at a value smaller than 1. This confirms our assumption that the structure of influences is evolving smoothly. Moreover, it drops to zero when $\delta$ increases, since there is no information flow in any windows.

Finally, by manipulating the matrix $I(X,Y)$, it is possible to probe a more detailed structure of the influences, introducing the information directionality flow
\begin{equation}
\Delta_n = \sum_{j \neq n} I(X_n, X_j) - \sum_{i \neq n} I(X_i, X_n) ~.
\label{eq:Delta_n}
\end{equation}
The first term is a summation across columns, measuring the total information flow originating from the stock $n$, while the second is a summation across rows, measuring the total information flow directed toward $n$, originated from other stocks. The difference between these two terms provides a measure of whether the stock $n$ is influencing the market ($\Delta_n > 0$) or is being influenced by the market ($\Delta_n < 0$), and by how much. A plot of the stock's directionality measures with time provides knowledge about how the stock's role in the market has been evolving, as shown in Figs.~3, 4 and 5.
This picture also provides another check on the slow evolution of the matrices $I(X_i, X_j)$ during several time windows.

\section*{Discussion}
\label{sec:discussion}
To validate the results obtained using a different approach we employed another method to investigate the data.
This method is based on extracting all the pairs $\{X,Y\}$ for which $T_{X\rightarrow Y}$ is larger than a given threshold and then, using the same threshold, extracting all pairs of a \emph{surrogate dataset} of time series for which the same condition is satisfied.
Finally, we compare the number of links extracted in the two cases.
The surrogate dataset, obtained following the protocol outlined above, does not include genuine interactions between its degrees of freedom at any of the time windows considered and we do not expect it to reproduce the patterns found in Fig.~1. More specifically, we do not expect to observe an abrupt increase in the number of links around crises.
We compute the ratio of the number of directed links detectable in the real and surrogate datasets and we denote this quantity by $1/r_t$, in order to be consistent with the notation used in the Supporting Information, see eq. (\ref{eq:SIdefrx}).
Figure~6 shows that such significant increases are not observed for the surrogate data and therefore these are genuine phenomena of the real dataset.

We would like to point out that while we mainly discuss results obtained for $k=2$, we also carried out the analysis for  larger $k$ values. Computing the matrices $T_{Y\rightarrow X}$ for larger $k$ values requires much more time; additionally we were unsuccessful in extracting meaningful information $I(X,Y)$ even for $k=3$.
This is disappointing but is confirmed by other works~\cite{marschinski2002analysing} studying the TE measure (\ref{eq:tildeTYX}) by varying the parameter $k$, where it has been shown that large values of $k$ provide uninformative results.
A possible reason is the signal-to-noise ratio, which is very small already for $k=2$ and is presumably lost in more complicated models of influences. Moreover, when increasing $k$ at a fixed $\Omega$, we notice that the quality of the probability estimation in Eq.~(\ref{eq:tildeTYX}) decreases.

Finally we repeated the analysis presented in the previous section by using the row prices rather than the returns, even if they are usually not studied in this context.
As explained in the Supporting information, there are more or less conservative link extraction protocols, where, roughly speaking, the first ones lead to extract (maybe too) few trustworthy links and the last ones lead to extract (maybe too) many links at the price of considering many spurious links.
Employing a less conservative protocol when considering time series of prices we obtained results which resemble those presented above while the exact same protocol used before leads to uninteresting results where no information can be extracted in any window, as shown in Fig.~7.
This is consistent with the common belief that returns, in general, carry more useful information than prices.

\begin{figure}[ht]
\centering
\includegraphics[width=130mm]{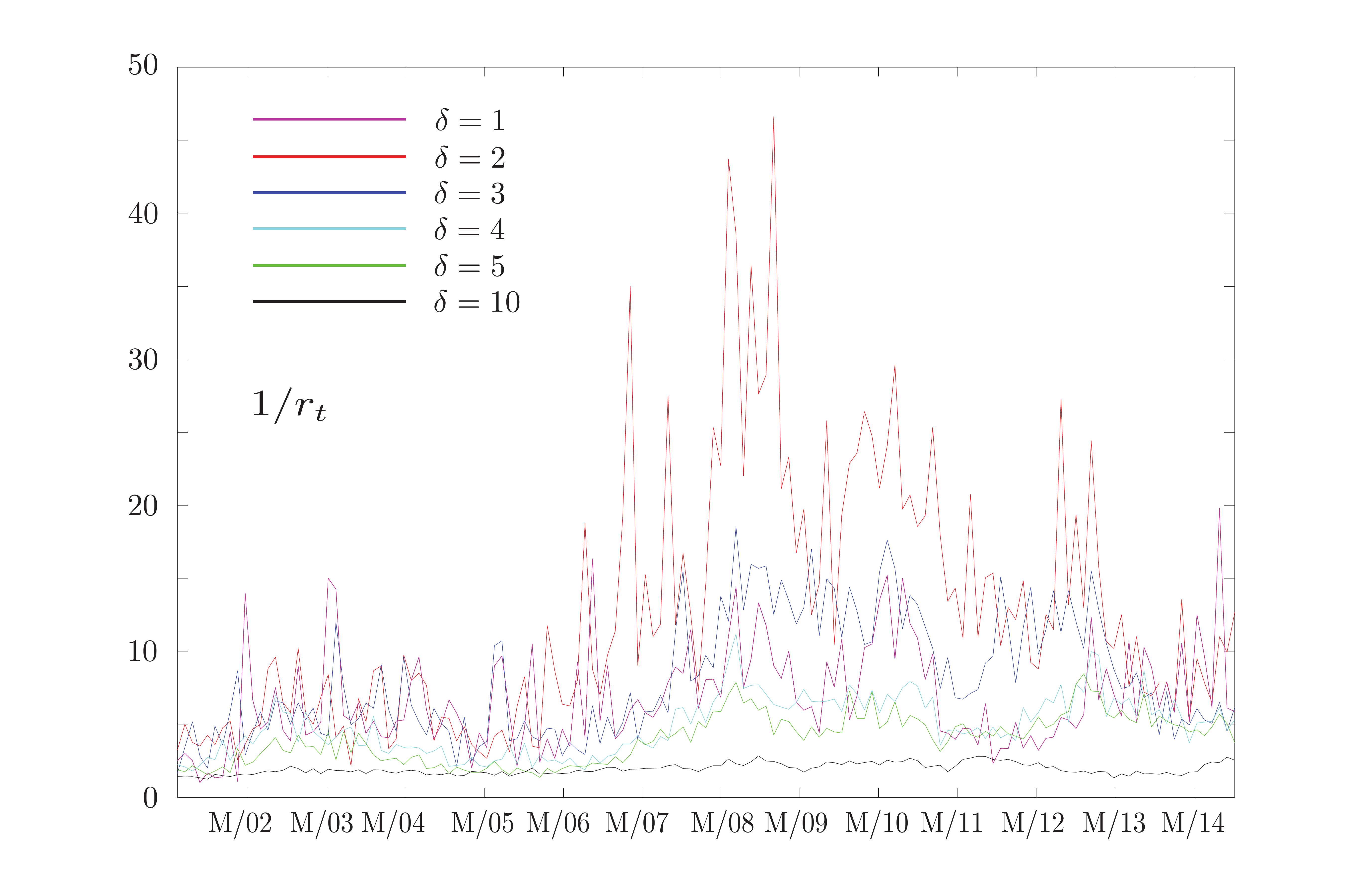}
\label{Fig6}
\caption{We denote by $1/r_t$ the ratio of the number of directed links detectable in the real and surrogate datasets. This is made in order to be consistent with the notation used in the The threshold used is $0.03$ but similar qualitative results are obtained for other values. We see that as $\delta$ increases this ratio approaches $1$, while at short time scales it resembles the results of Fig.~1.}
\end{figure}

\begin{figure}[ht]
\centering
\includegraphics[width=130mm]{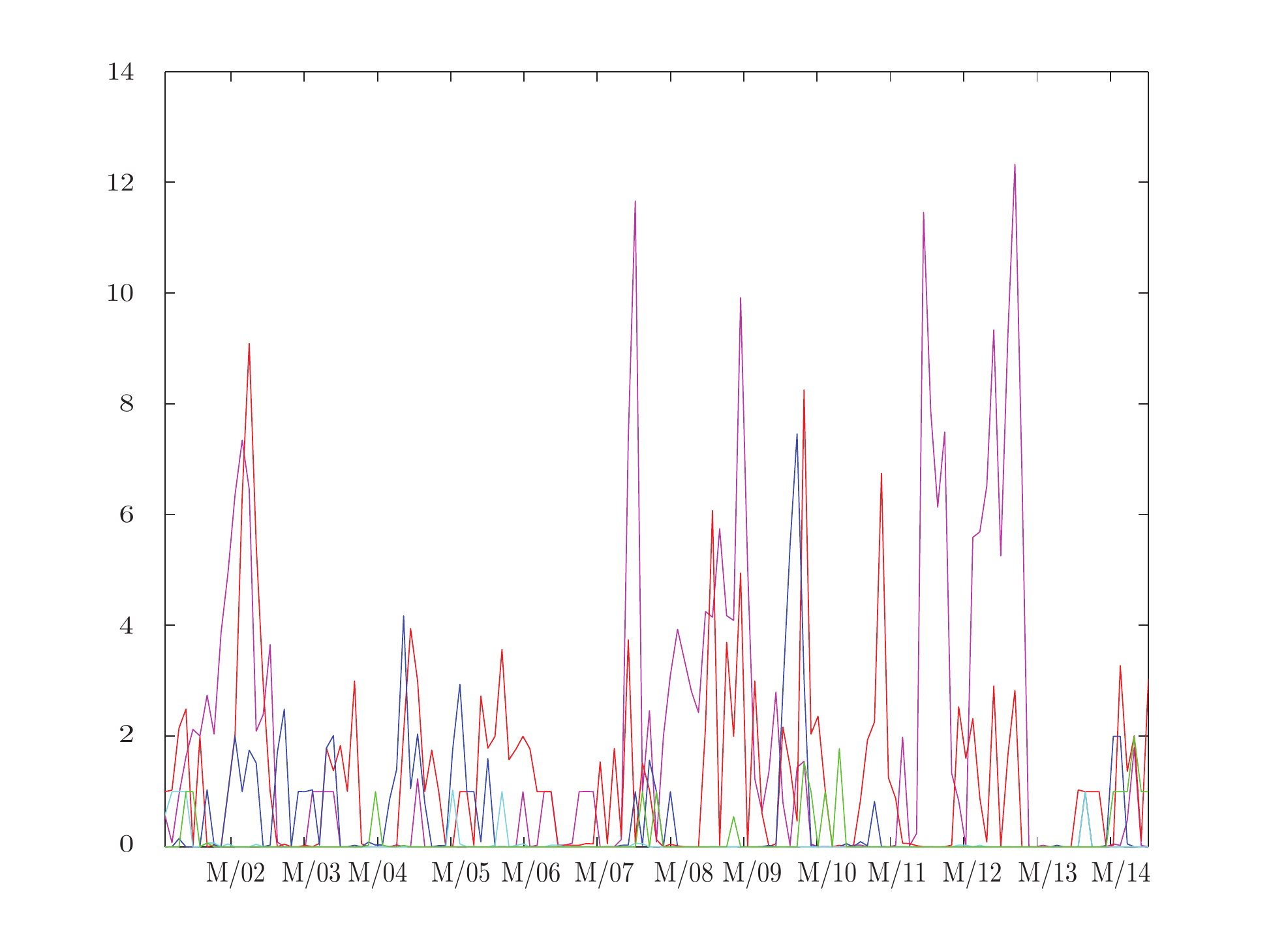}
\label{Fig7}
\caption{Behavior of the total information flow  $\mathcal{I}(w)$, defined in eq. (\ref{eq:totalIw}), at different time differences $\delta$, when computed for daily prices rather than returns. Each time window $w$ is $500$ days long. The date associated with each $w$ is the middle of the time window considered. The $x$-axis tick marks correspond to the dates of the first of March each year. No meaningful information can be obtained from this analysis in contrast to the results presented in Fig.~1.}
\end{figure}

\section*{Conclusions}

In this work we develop an information theoretical measure to compute the information flow among time series.
The new method improves on the standard linear methods such as retarded correlations or the Granger causality measure~\cite{fiedor2014information, souza2016nonlinear}, allowing one to extract influences that would be otherwise hardly detectable.
Our analysis describes the structure of the market in a long time period spanning from $2000$ to $2015$, exposing a critical behavior at times of financial crises.
The analysis supports the dominant viewpoint according to which no information can be extracted at long time scales (of days) at normal day-to-day operations, but sheds light on the emergence of interlinked information flow at times that are close critical events. Near such events the corresponding financial time series, may be very different from random walks and exhibit information flows at the time scales of days, as observed in~\cite{fiedor2015}.
We clearly observe that information flow between stocks during non-crises periods have short-lived effects on the market, whereas during crises they exert their influence over a larger range of time scales, having longer-lasting effects.
Interestingly, we notice that early signals of crashes could already be detected in the markets months before the full manifestation of the crises. The Lehman Brothers bankruptcy, that arguably marks the onset of the 2008 financial crisis, is dated to September 15, 2008, and while this event is clearly recognizable in Fig.~1, we also notice that the cross-stock information flow started to grow months in advance, when the American subprime mortgage market started to unfold.
Whether or not this method can predict catastrophic events is unclear, but it can definitely measure the susceptibility of financial markets and their robustness to volatility as it exposes the strengthening of long-range correlations, in analogy to complex systems close to a phase transition.
Future work will apply this method to probe information flows at shorter time scales, which represent every-day operation. Looking at finer time scale would provide useful insights in order to characterize the state of markets even far from global crises.

\section*{Supporting Information}

\subsection*{Transfer Entropy}
In this section we present the information theoretical tools at the heart of the transfer entropy measure defined in Eq.~(\ref{eq:tildeT}).
Let us consider two time series, $X(t)$ and $Y(t)$, and denote by $x^{(k)}_t$ the $k$ past time steps of the time series $X$ at time $t$: $\{x_{t-k}, \ldots, x_{t-1} \}$.
The difference between the probability of observing $x_{t}$ given $x^{(k)}_t$ and the probability of observing $x_{t}$ given $x^{(k)}_t$ and $y^{(l)}_t$ can be computed by using the Kullback-Leibler divergence:
\begin{equation}
\label{eq:KL_X_Y}
D_{Y\rightarrow X} \left(x^{(k)}_t, y^{(l)}_t \right) = \sum_{x_{t} } p\left(x_{t} | x^{(k)}_t, y^{(l)}_t \right) \log_2 \left( \frac{p\left(x_{t} | x^{(k)}_t, y^{(l)}_t \right)}{p\left(x_{t} | x^{(k)}_t \right)}  \right)\:.
\end{equation}
This object is zero if $Y(t)$ contains no information about $x_{t}$ and positive otherwise.
The transfer entropy from $Y(t)$ to $X(t)$, denoted by $T_{Y\rightarrow X}$, is the average over the past observations of~(\ref{eq:KL_X_Y}):
\begin{equation}
T^{(TE)}_{Y\rightarrow X} = \mathds{E}_{\{x^{(k)}_t, y^{(l)}_t \}} \left[ D_{Y\rightarrow X} \left(x^{(k)}_t, y^{(l)}_t \right) \right]\:~.
\label{eq:TEYtoX}
\end{equation}
It accounts for the information gained about the present value of the time series $X(t)$ by also considering the $l$ past values of the time series $Y(t)$, in addition to the $k$ past values of $X(t)$.
The time steps scale can be generalized from $1$ to a general value $\delta$. In this case we have $x_{t}^{(k)}=\{ x_{t-k \: \delta}, \ldots, x_{t-\delta}\}$.
Usually, the computation of TE is done by setting $k=l=1$ for computational reasons; moreover, increasing $k$ may destroy meaningful information flow, as shown in~\cite{marschinski2002analysing}.

The computation of TE from the observed time series requires estimation of the various probability distributions in Eq.~(\ref{eq:TEYtoX}). Among the proposed estimation methods is STE~\cite{staniek2008symbolic}, which employs the technique of symbolization.
A $k-$dimensional symbol of the time series $X(t)$ at time $t$ is defined by ordering the values $x_{t+\delta}^{(k)}=\{ x_{t-(k-1)\delta}, \ldots, x_{t-\delta}, x_{t} \}$ in an ascending order. The symbol associated with this part of the time series is denoted by $\hat{x}^{k}_{t+\delta}$. More details on this protocol are given in the next section.
The symbolic transfer entropy is then defined by
\begin{equation}
T^{(STE)}_{Y\rightarrow X} = \sum_{\hat{x}^{k}_{t+\delta}, \hat{x}^{k}_t, \hat{y}^{k}_t } p\left(\hat{x}^{k}_{t+\delta}, \hat{x}^{k}_t, \hat{y}^{k}_t \right) \log_2 \left( \frac{p\left(\hat{x}^{k}_{t+\delta} | \hat{x}^{k}_t, \hat{y}^{k}_t \right)}{p\left(\hat{x}^{k}_{t+\delta} | \hat{x}^{k}_t \right)}  \right)\:,
\label{eq:tildeTYX}
\end{equation}
which directly follows from Eq.~(\ref{eq:TEYtoX}) once that explicit values are replaced by symbols.
Assuming stationarity, the required probability distributions can be estimated by computing the occurrences of symbols in the time series, suppressing the effect of noise and bypassing the fine-tuning of parameters in probability distribution estimation protocols. Notice that each symbol is drawn from the values of the time series at $k$ time steps into the past and so that a single symbol contains information from $k$ historic time steps.

The measure we introduced in Eq.~(\ref{eq:tildeT}) is very similar to STE but rather than dealing with $k$-dimensional symbols, it aims to predict $k+1$-dimensional symbols from $k$-dimensional ones, but with a modest computational cost.
The main reason to introduce this measure has been to gain computational power in predicting symbols of $k+1$ literals; this was particularly important due to the long preprocessing time required for the type of datasets analyzed.
The reason is that for each pair of stocks there must be a one-to-one correspondence between the respective trading days. Days when one of the two is not traded are potentially problematic since they may shift the time index in one of the two and interfere with the causality relations.
To deal with this issue we adopted a practical approach by removing all the non-common days in each pair of the time series considered.
Since the number of disregarded days in each pair of stocks does not exceed 10 days this may seems a minor difficulty.
By the way it requires a larger pre-processing effort, since we cannot symbolize the time series once and for all before computing the matrix $T$.
Instead, we have to pre-process the time series of the stocks on a pair-by-pair basis before symbolizing it for each pair in a dedicated manner, which slows down the process considerably.

\subsection*{Symbolization}
Here we provide further details on the symbolization technique.
A $k-$dimensional symbol of the time series $X(t)$ at time $t$,
\begin{equation}
\hat{x}^{k,\delta}_{t+\delta}=\{ j_1, \ldots, j_{k-1}, j_k \}\:,
\end{equation}
is defined by ordering the values $x_{t+\delta}^{(k)}=\{ x_{t-(k-1)\delta}, \ldots, x_{t-\delta}, x_{t} \}$ in an ascending order $\{ x_{t-(j_1-1)\delta}, \ldots, x_{t-(j_{k-1}-1)\delta}, x_{t-(j_k-1)\delta} \}$.
If there are repeated values, the one with the smaller index comes first~\cite{staniek2008symbolic}.
Here we are going to give a few examples, making the dependance on the time steps scale $\delta$ explicit.
Let us consider the time series in Table~\ref{tabtable0}.
\begin{table}[!ht]
\label{tabsequence}
\centering
\caption{A time series $X(t)$ and the respective time index $t$. The symbols constructed from this time series are provided in Table~\ref{tabtable1}.}
\begin{tabular}{|c|r|r|r|r|r|r|r|r|r|r|r|r|}
\hline
$X(t)$ & 13 & 22 & 45 & 60 & 12 & 33 & 70 & 19 & 20 & 15 & 12 & 42 \\ \hline
$t$ & 1 & 2 & 3 & 4 & 5 & 6 & 7 & 8 & 9 & 10 & 11 & 12 \\ \hline
\end{tabular}
\label{tabtable0}
\end{table}

Following the definition given above, and in a way of demonstration, we provide some of the symbols constructed from this time series in Table~\ref{tabtable1}.
\begin{table}[!ht]
\centering
\caption{
This table contain samples of symbols extracted from the sequence in Table \ref{tabtable0}. The first three symbols have $k=2$, the last three have $k=3$. For each case we evaluate three different time scales $\delta=1,2,3$.}
\begin{tabular}{|c|c|c|c|}
\hline
\multicolumn{4}{| c |}{\bf Symbolization} \\ \thickhline
\hline
 & Symbol & Sequence considered & Symbol value \\ \hline
\multirow{3}{*}{$k=2$}
& $\hat{x}^{2,1}_{12}$ & $\{15,12\}$ & $\{2, 1\}$ \\
& $\hat{x}^{2,2}_{11}$ & $\{70, 20\}$ & $\{2, 1\}$ \\
& $\hat{x}^{2,3}_{10}$ & $\{60, 70\}$ & $\{1, 2\}$ \\ \hline
\multirow{3}{*}{$k=3$}
& $\hat{x}^{3,1}_{12}$ & $\{20,15,12\}$ & $\{3, 2, 1\}$ \\
& $\hat{x}^{3,2}_{11}$ & $\{12, 70, 20\}$ & $\{1, 3, 2\}$ \\
& $\hat{x}^{3,3}_{10}$ & $\{13, 60, 70\}$ & $\{1, 2, 3\}$ \\ \hline
\end{tabular}
\label{tabtable1}
\end{table}
\subsection*{Details on the evaluation of $I(X,Y)$}
\begin{figure}[ht]
\centering
\includegraphics[width=130mm]{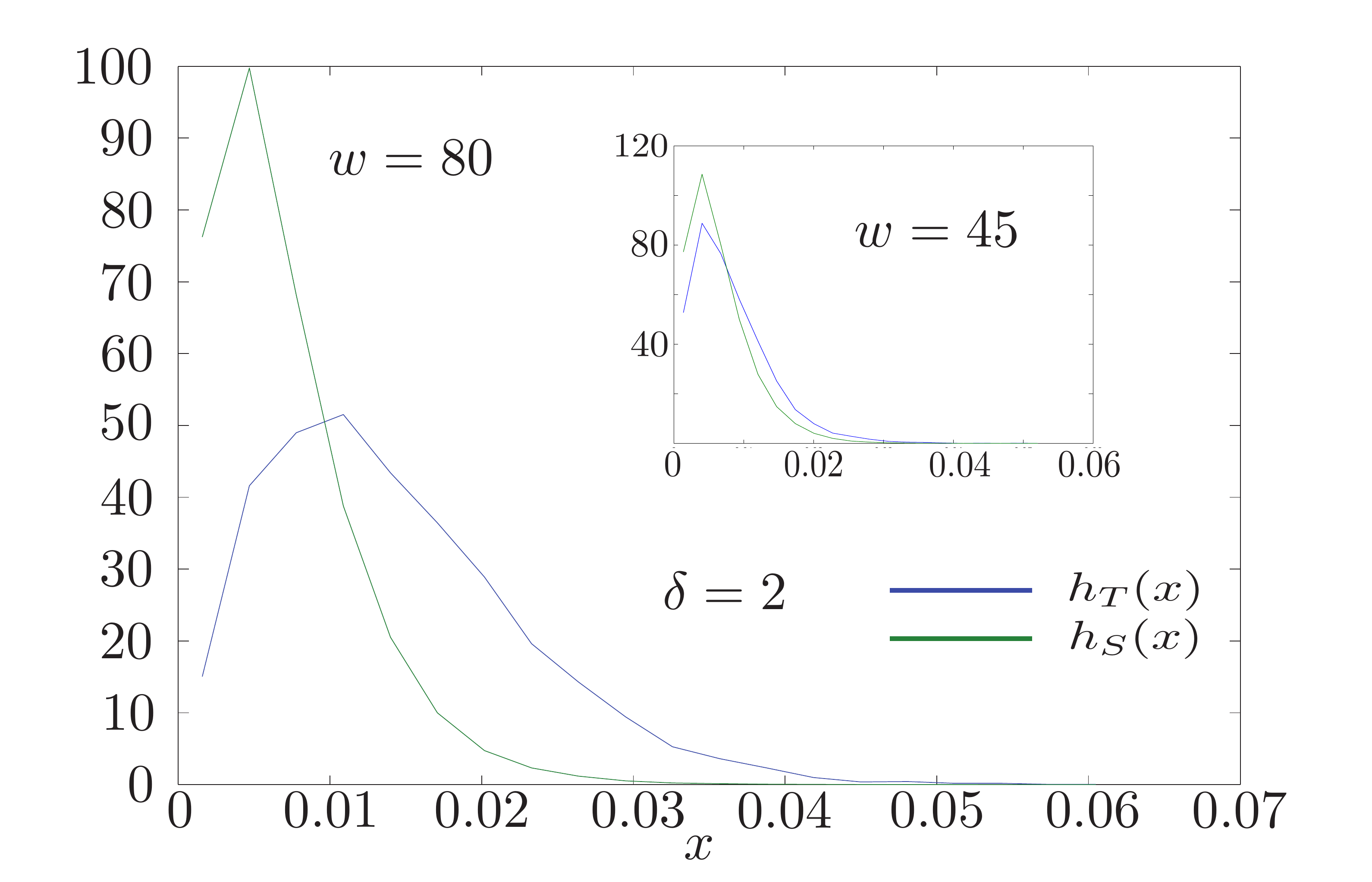}
\label{Fig8}
\caption{Transfer entropy values - real and surrogate data. Histograms of values found in the sets $\mathcal{T}$ and $\mathcal{S}$ for $w=80$, i.e. the period of November 2008, using the time scale of Fig.~1. The inset shows the same quantities computed at $w=45$, i.e. September 2005. While in the second case no information flows can be detected, in the first, using the protocol discussed in this section, many directed influenced can be obtained. This matrix values refer to $\delta=2$, for which the amount of information is maximized.}
\end{figure}
\begin{figure}[ht]
\centering
\includegraphics[width=130mm]{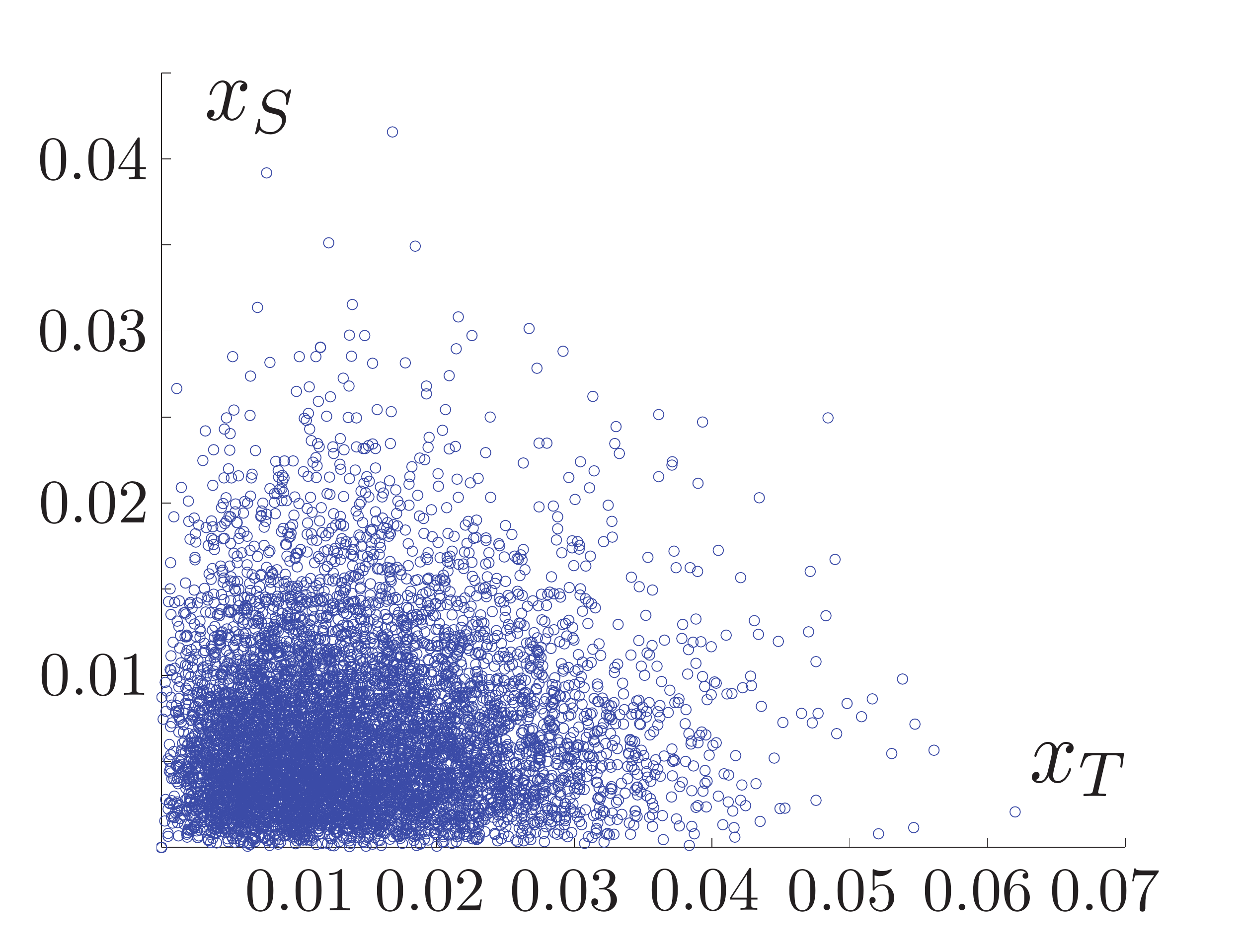}
\label{Fig9}
\caption{Pplot of the values found in $\mathcal{T}$ at $\delta=2$ in $w=80$ versus those found in the surrogate dataset at the same $w$ and $\delta$. The values do not appear to show any correlation between the two.}
\end{figure}
In this section we provide further details on the method used to evaluate the values $I(X,Y)$, used in Eqs.~(\ref{eq:eqDdiff}), (\ref{eq:totalIw}) and (\ref{eq:Delta_n}).
These quantities reflect the amount of genuine information flow from time series $X(t)$ to the time series $Y(t)$ and are obtained by processing the measure introduced in Eq.~(\ref{eq:tildeT}).
Cleaning these matrices from spurious values is not an easy task; after the construction of a null model, one needs to employ a thresholding method to filter out random effects.

For each window $w$, we consider the set $\mathcal{T}$ of TEs from the true dataset and we formed a benchmark set $\mathcal{S}$ of TEs by collecting the values obtained from the surrogate datasets in the $21$ windows bracketing $w$, i.e. $\{w-10, \ldots, w , \ldots, w+10 \}$.
In other words, we assumed that the null models of consecutive time windows do not differ too much, given that the two windows are shifted by $25$ days, corresponding to $5\%$ of their length.
A comparison of the histograms $h_T(x)$ and $h_S(x)$ of the set $\mathcal{T}$ and $\mathcal{S}$ gives a crude estimations of the $p$-values of the TEs computed for the true dataset, i.e. of the probability that the values $x=T_{X \rightarrow Y}$ obtained for the true dataset has been obtained at random.
This can be done computing, for each $x$, the ratio
\begin{equation}
r(x)=\frac{\int_{x}^{\infty}dx'\: h_S(x')}{\int_{x}^{\infty}dx'\: h_T(x')}\:.
\label{eq:SIdefrx}
\end{equation}
The ratio $r$ decreases to $0$ as $x$ increases: small $r$ values are associated with $x$ values for which it is more likely to have a genuine information flow.
Thus, we associated a weigh to each pair $\{X \rightarrow Y\}$ given by
\begin{equation}
I(X,Y)=\frac{1}{e^{2 a(r(x)-r^*)}+1}
\end{equation}
where $x=T_{X \rightarrow Y}$, $a=100$ and $r^*=0.03$.
These two histograms can be seen in Fig.~8 for two particular time windows.
One of the possible pitfalls of this method is that values in $\mathcal{T}$ are correlated to values in $\mathcal{S}$. If this were to be the case, we would underestimate the number of detected influences; however, as the scatterplot in Fig.~9 shows, this is not the case.
Using this value of $r^*$ can be seen as a soft thresholding method which roughly corresponds to considering a $p$-value smaller than $0.05$.
Statistically validated networks are obtained by considering much smaller thresholds that take into account multiple comparison effects. 
By the way, employing such a strict protocol would provide poor results in the present case because the possible retrievable information is very small and we are forced to adopt a less conservative protocol.
Nevertheless, our results have been cross-validated by using the method described in the Discussion section, and the fact that the null model is unable to reproduce the total information flow patterns detected in the original dataset validates the results of the analysis.

\section*{Acknowledgments}
We would like to thank David Lowe for interesting discussions. 
This work is supported by The Leverhulme Trust grant RPG-2013-48. E.T. also wishes to thank the Lee Hysan Foundation of Hong Kong for funding.


%
%
%

\bibliographystyle{plos2015}

\begin{thebibliography}{10}

\bibitem{mantegna2000introduction}
Mantegna RN, Stanley HE.
\newblock An introduction to econophysics: correlation and complexity in
  finance.
\newblock Cambridge, UK: Cambridge University. 2000.

\bibitem{plerou2000econophysics}
Plerou V, Gopikrishnan P, Rosenow B, Amaral LA, Stanley HE.
\newblock Econophysics: financial time series from a statistical physics point
  of view.
\newblock Physica A: Statistical Mechanics and its Applications.
  2000;279(1):443--456.

\bibitem{bouchaud2003theory}
Bouchaud JP, Potters M.
\newblock Theory of financial risk and derivative pricing: from statistical
  physics to risk management.
\newblock Cambridge university press; 2003.

\bibitem{plerou1999universal}
Plerou V, Gopikrishnan P, Rosenow B, Amaral LAN, Stanley HE.
\newblock Universal and nonuniversal properties of cross correlations in
  financial time series.
\newblock Physical Review Letters. 1999;83(7):1471.

\bibitem{laloux2000random}
Laloux L, Cizeau P, Potters M, Bouchaud JP.
\newblock Random matrix theory and financial correlations.
\newblock International Journal of Theoretical and Applied Finance.
  2000;3(03):391--397.

\bibitem{fenn2011temporal}
Fenn DJ, Porter MA, Williams S, McDonald M, Johnson NF, Jones NS.
\newblock Temporal evolution of financial-market correlations.
\newblock Physical review E. 2011;84(2):026109.

\bibitem{sornette2012dragon}
Sornette D, Ouillon G.
\newblock Dragon-kings: mechanisms, statistical methods and empirical evidence.
\newblock The European Physical Journal Special Topics. 2012;205(1):1--26.

\bibitem{gualdi2015tipping}
Gualdi S, Tarzia M, Zamponi F, Bouchaud JP.
\newblock Tipping points in macroeconomic agent-based models.
\newblock Journal of Economic Dynamics and Control. 2015;50:29--61.

\bibitem{mantegna1999hierarchical}
Mantegna RN.
\newblock Hierarchical structure in financial markets.
\newblock The European Physical Journal B-Condensed Matter and Complex Systems.
  1999;11(1):193--197.

\bibitem{tumminello2005tool}
Tumminello M, Aste T, Di~Matteo T, Mantegna RN.
\newblock A tool for filtering information in complex systems.
\newblock Proceedings of the National Academy of Sciences of the United States
  of America. 2005;102(30):10421--10426.

\bibitem{allen2008networks}
Allen F, Babus A.
\newblock Networks in finance.
\newblock Wharton Financial Institutions Center Working Paper; 2008.

\bibitem{haldane2013rethinking}
Haldane AG.
\newblock Rethinking the financial network.
\newblock In: Fragile stabilit{\"a}t--stabile fragilit{\"a}t. Springer; 2013.
  p. 243--278.

\bibitem{strogatz2001exploring}
Strogatz SH.
\newblock Exploring complex networks.
\newblock Nature. 2001;410(6825):268--276.

\bibitem{elton1971improved}
Elton EJ, Gruber MJ.
\newblock Improved forecasting through the design of homogeneous groups.
\newblock The Journal of Business. 1971;44(4):432--450.

\bibitem{wilson1971renormalization}
Wilson KG.
\newblock Renormalization group and critical phenomena. I. Renormalization
  group and the Kadanoff scaling picture.
\newblock Physical review B. 1971;4(9):3174.

\bibitem{sornette1998discrete}
Sornette D.
\newblock Discrete-scale invariance and complex dimensions.
\newblock Physics reports. 1998;297(5):239--270.

\bibitem{sornette2001significance}
Sornette D, Johansen A, et~al.
\newblock Significance of log-periodic precursors to financial crashes.
\newblock Quantitative Finance. 2001;1(4):452--471.

\bibitem{dakos2008slowing}
Dakos V, Scheffer M, van Nes EH, Brovkin V, Petoukhov V, Held H.
\newblock Slowing down as an early warning signal for abrupt climate change.
\newblock Proceedings of the National Academy of Sciences.
  2008;105(38):14308--14312.

\bibitem{scheffer2009early}
Scheffer M, Bascompte J, Brock WA, Brovkin V, Carpenter SR, Dakos V, et~al.
\newblock Early-warning signals for critical transitions.
\newblock Nature. 2009;461(7260):53--59.

\bibitem{scheffer2012anticipating}
Scheffer M, Carpenter SR, Lenton TM, Bascompte J, Brock W, Dakos V, et~al.
\newblock Anticipating critical transitions.
\newblock science. 2012;338(6105):344--348.

\bibitem{bonanno2003topology}
Bonanno G, Caldarelli G, Lillo F, Mantegna RN.
\newblock Topology of correlation-based minimal spanning trees in real and
  model markets.
\newblock Physical Review E. 2003;68(4):046130.

\bibitem{onnela2003dynamics}
Onnela JP, Chakraborti A, Kaski K, Kertesz J, Kanto A.
\newblock Dynamics of market correlations: Taxonomy and portfolio analysis.
\newblock Physical Review E. 2003;68(5):056110.

\bibitem{bonanno2004networks}
Bonanno G, Caldarelli G, Lillo F, Miccich{\`e} S, Vandewalle N, Mantegna RN.
\newblock Networks of equities in financial markets.
\newblock The European Physical Journal B-Condensed Matter and Complex Systems.
  2004;38(2):363--371.

\bibitem{tumminello2007correlation}
Tumminello M, Di~Matteo T, Aste T, Mantegna R.
\newblock Correlation based networks of equity returns sampled at different
  time horizons.
\newblock The European Physical Journal B. 2007;55(2):209--217.

\bibitem{granger1988some}
Granger CW.
\newblock Some recent development in a concept of causality.
\newblock Journal of econometrics. 1988;39(1):199--211.

\bibitem{curme2015emergence}
Curme C, Tumminello M, Mantegna RN, Stanley HE, Kenett DY.
\newblock Emergence of statistically validated financial intraday lead-lag
  relationships.
\newblock Quantitative Finance. 2015;15(8):1375--1386.

\bibitem{huth2014high}
Huth N, Abergel F.
\newblock High frequency lead/lag relationships-Empirical facts.
\newblock Journal of Empirical Finance. 2014;26:41--58.

\bibitem{schreiber2000measuring}
Schreiber T.
\newblock Measuring information transfer.
\newblock Physical review letters. 2000;85(2):461.

\bibitem{fiedor2014information}
Fiedor P.
\newblock Information-theoretic approach to lead-lag effect on financial
  markets.
\newblock The European Physical Journal B. 2014;87(8):1--9.

\bibitem{fiedor2015}
Fiedor P.
\newblock Granger-causal nonlinear financial networks.
\newblock Journal of Network Theory in Finance. 2015;1(2):53--82.

\bibitem{fama1965behavior}
Fama EF.
\newblock The behavior of stock-market prices.
\newblock The journal of Business. 1965;38(1):34--105.

\bibitem{bossomaier2013information}
Bossomaier T, Barnett L, Harr{\'e} M.
\newblock Information and phase transitions in socio-economic systems.
\newblock Complex Adaptive Systems Modeling. 2013;1(1):1.

\bibitem{fiedor2014frequency}
Fiedor P.
\newblock Frequency effects on predictability of stock returns.
\newblock In: 2014 IEEE Conference on Computational Intelligence for Financial
  Engineering \& Economics (CIFEr). IEEE; 2014. p. 247--254.

\bibitem{staniek2008symbolic}
Staniek M, Lehnertz K.
\newblock Symbolic transfer entropy.
\newblock Physical Review Letters. 2008;100(15):158101.

\bibitem{rubinov2010complex}
Rubinov M, Sporns O.
\newblock Complex network measures of brain connectivity: uses and
  interpretations.
\newblock Neuroimage. 2010;52(3):1059--1069.

\bibitem{vicente2011transfer}
Vicente R, Wibral M, Lindner M, Pipa G.
\newblock Transfer entropy-a model-free measure of effective connectivity for
  the neurosciences.
\newblock Journal of computational neuroscience. 2011;30(1):45--67.

\bibitem{borge2016dynamics}
Borge-Holthoefer J, Perra N, Gon{\c{c}}alves B, Gonz{\'a}lez-Bail{\'o}n S,
  Arenas A, Moreno Y, et~al.
\newblock The dynamics of information-driven coordination phenomena: A transfer
  entropy analysis.
\newblock Science advances. 2016;2(4):e1501158.

\bibitem{souza2016nonlinear}
Souza TT, Aste T.
\newblock A nonlinear impact: evidences of causal effects of social media on
  market prices.
\newblock arXiv preprint arXiv:160104535. 2016.

\bibitem{marschinski2002analysing}
Marschinski R, Kantz H.
\newblock Analysing the information flow between financial time series.
\newblock The European Physical Journal B-Condensed Matter and Complex Systems.
  2002;30(2):275--281.

\bibitem{kwon2008information}
Kwon O, Yang JS.
\newblock Information flow between stock indices.
\newblock EPL (Europhysics Letters). 2008;82(6):68003.

\bibitem{YahooFin}
{Yahoo! Finance} website.
\newblock \url{http://finance.yahoo.com}.

\bibitem{theiler1996constrained}
Theiler J, Prichard D.
\newblock Constrained-realization Monte-Carlo method for hypothesis testing.
\newblock Physica D: Nonlinear Phenomena. 1996;94(4):221--235.

\bibitem{schreiber1998constrained}
Schreiber T.
\newblock Constrained randomization of time series data.
\newblock Physical Review Letters. 1998;80(10):2105.

\bibitem{theiler1992testing}
Theiler J, Eubank S, Longtin A, Galdrikian B, Farmer JD.
\newblock Testing for nonlinearity in time series: the method of surrogate
  data.
\newblock Physica D: Nonlinear Phenomena. 1992;58(1-4):77--94.



\end{thebibliography}

\end{document}